\documentclass[pra,twocolumn,twoside,showpacs]{revtex4}

\usepackage{graphicx}
\usepackage{amsmath,bm}
\usepackage{ae}

\newcommand{\eq}[1]{\begin{equation}#1\end{equation}}
\newcommand{\eqmulti}[1]{\begin{equation}\begin{split}#1\end{split}\end{equation}}

\newcommand{\ket}[1]{\ensuremath{\,|{#1}\rangle}}

\newcommand{\matrixe}[3]{\ensuremath{\langle{#1}|\,{#2}\,|{#3}\rangle}}
\newcommand{\expect}[1]{\ensuremath{\langle{#1}\rangle}}
\newcommand{\op}[1]{\ensuremath{\bm{\mathrm{#1}}}}
\newcommand{\adj}[1]{\ensuremath{{{#1}}^{\dag}}}

\newcommand{\aO}{\ensuremath{\op{a}}}
\newcommand{\aaO}{\ensuremath{\adj{\op{a}}}}
\newcommand{\cO}{\ensuremath{\op{c}}}
\newcommand{\ccO}{\ensuremath{\adj{\op{c}}}}
\newcommand{\nO}{\ensuremath{\op{n}}}

\newcommand{\AO}{\ensuremath{\op{A}}}
\newcommand{\AAO}{\ensuremath{\adj{\op{A}}}}
\newcommand{\HO}{\ensuremath{\op{H}}}
\newcommand{\JO}{\ensuremath{\op{J}}}
\newcommand{\TO}{\ensuremath{\op{T}}}

\newcommand{\eV}{\ensuremath{\vec{e}}}
\newcommand{\vV}{\ensuremath{\vec{v}}}
\newcommand{\xV}{\ensuremath{\vec{x}}}
\newcommand{\yV}{\ensuremath{\vec{y}}}

\newcommand{\nablaV}{\ensuremath{\vec{\nabla}}}

\newcommand{\ii}{\ensuremath{\mathrm{i}}}
\newcommand{\ee}{\ensuremath{\mathrm{e}}}

\newcommand{\SF}{\ensuremath{\textrm{s}}}

\begin{document}

\title{Superfluidity and Interference Pattern of Ultracold Bosons in Optical Lattices}

\author{R. Roth}
\author{K. Burnett}

\affiliation{Clarendon Laboratory, University of Oxford,
  Parks Road, Oxford OX1 3PU, United Kingdom}

\date{\today}

\begin{abstract} 
  We present a study of the superfluid properties of atomic Bose gases in 
  optical lattice potentials using the Bose-Hubbard model. To do this, we
  use a microscopic definition of the superfluid fraction based on the
  response of the system to a phase variation imposed by means of twisted
  boundary conditions. We compare the superfluid fraction to other
  physical quantities, i.e., the interference pattern after ballistic
  expansion, the quasi-momentum distribution, and number fluctuations. We
  have performed exact numerical calculations of all these quantities for
  small one-dimensional systems. We show that the superfluid fraction
  alone exhibits a clear signature of the Mott-insulator transition.
  Observables like the fringe visibility, which probe only ground state
  properties, do not provide direct information on superfluidity and the
  Mott-insulator transition.
\end{abstract}

\pacs{03.75.Fi, 05.30.Jp, 73.43.Nq, 67.40.-w}

\maketitle


Ultracold atomic gases in optical lattices provide a unique framework for
the experimental study of fundamental quantum phenomena in interacting
many-body systems. This is especially true for the exploration of quantum
phase transitions such as the superfluid to Mott-insulator transition
observed in a recent pioneering experiment \cite{GrMa02}. The remarkable
degree of experimental control over all the relevant parameters---density,
interaction strength, lattice geometry and dimensionality---allows much
more detailed studies of the complicated mechanisms behind quantum phase
transition than conventional solid state systems. 

It is clearly the case that the most important quantity for characterizing
the superfluid-to-insulator transition is the superfluid density or
superfluid fraction $f_{\SF}$. The aim of this paper is to set up the
general theoretical framework for the calculation of the superfluid
fraction within the Bose-Hubbard model and to compare it, on the formal
level, with quantities being measured at the moment. These include the
interference pattern after ballistic expansion, the quasi-momentum
distribution as well as the number fluctuations which are important in
applications. We perform exact numerical calculations for the
Mott-insulator transition in an one-dimensional system to demonstrate the
relationship, and more importantly the differences, between the superfluid
fraction and various ground state observables, most notably the visibility
of the interference pattern.

\enlargethispage{2ex}

\paragraph*{Superfluidity.}

The concept of superfluidity is closely related to the existence of a 
condensate in the interacting many-boson system \cite{Legg99}. Formally,
the one-body density matrix $\rho^{(1)}(\xV,\xV')$ has to have exactly one
macroscopic eigenvalue which defines the number of particles $N_{0}$ in
the condensate; the corresponding eigenvector describes the condensate wave
function $\phi_0(\xV) = \ee^{\ii\theta(\xV)} |\phi_0(\xV)|$. A spatially
varying condensate phase $\theta(\xV)$ is associated with a velocity field
for the condensate by
\eq{ \label{eq:velocity_sf}
  \vV_0(\xV) = \frac{\hbar}{m}\; \nablaV \theta(\xV) \;.
}
This irrotational velocity field is identified with velocity of
the superfluid flow $\vV_{\SF}(\xV)\equiv\vV_0(\xV)$ \cite{Legg99}
and enables us to derive an expression for the superfluid fraction
$f_{\SF}=N_{\SF}/N$. Consider a system with a finite linear dimension $L$
in $\eV_1$-direction and a ground state energy $E_0$ calculated with
periodic boundary conditions. Now we impose a linear phase variation
$\theta(\xV) = \Theta\, x_1/L$ with a total twist angle $\Theta$
over the length of the system in the $\eV_1$-direction. Technically,
this can be achieved by introducing twisted boundary conditions of
the form $\Psi(\xV_1,...,\xV_i+L \eV_1,...)=\ee^{\ii\,\Theta}\;
\Psi(\xV_1,...,\xV_i,...)$ with respect to all coordinates of
the many-body wave function. The resulting ground state energy
$E_{\Theta}$ will depend on the phase twist. For very small twist angles
$\Theta\ll\pi$ the energy difference $E_{\Theta}-E_{0}$ can be attributed
to the kinetic energy $T_{\SF}$ of the superflow generated by the phase
gradient. Thus
\eq{ 
  E_{\Theta} - E_{0} \overset{!}{=} T_{\SF}
   = \tfrac{1}{2}m N f_{\SF} \vV_{\SF}^2 \;, 
}
where $m N f_{\SF}$ is the total mass of the superfluid component.
Replacing the superfluid velocity $\vV_{\SF}$ with the phase gradient
according to Eq. \eqref{eq:velocity_sf} leads to a fundamental relation
for the superfluid fraction \cite{Krau91,FiBa73}
\eq{ \label{eq:sffraction_cont}
  f_{\SF}
  = \frac{2 m}{\hbar^2} \frac{L^2}{N} \; \frac{E_{\Theta}-E_{0}}{\Theta^2}
  \;.
}     
Hence the superfluid fraction can be interpreted as a measure for the
stiffness of the system under an imposed phase variation. This
demonstrates that superfluidity is not a static ground state property but
rather the response of the system to an external perturbation. We should
note that this definition of superfluidity does not tell anything about
the stability of the superfluid flow at finite velocities.


We now transfer these findings to a lattice system described in the
framework of the Bose-Hubbard model \cite{JaBr98,Krau91,RoBu02}. For
simplicity we restrict ourselves to a regular one-dimensional lattice
composed of $I$ sites described by the Bose-Hubbard Hamiltonian
\eq{ \label{eq:hamiltonian}
  \HO_0 
  = -J \sum_{i=1}^{I} (\aaO_i \aO_{i+1} + \text{h.a.}) 
   + \frac{V}{2} \sum_{i=1}^{I} \nO_i (\nO_i-1) \;, 
}
where $\aaO_i$ creates a boson in the lowest Wannier state localized at
site $i$ and $\nO_i = \aaO_i\aO_i$. The first term describes the hopping
or tunneling between adjacent sites with a tunneling strength $J$, the
second term characterizes the on-site two-body interaction with a strength
$V$ \cite{JaBr98,OoSt01}. The hopping between the first and the last site of
the lattice is included ($I+1 \hat{=} 1$), which corresponds to periodic
boundary conditions. The generalization to three-dimensional lattices is
straight forward.

In order to compute the energy of the system with an imposed phase twist
$\Theta$ we map the twisted boundary conditions by means of a local
unitary transformation onto the Hamiltonian \cite{ShSu90,Poil91}. This
leads to a twisted Hamiltonian of the form
\eq{ \label{eq:hamiltonian_twist}
  \HO_{\Theta}
  = -J \sum_{i=1}^{I} (\ee^{-\ii\, \Theta/I}\;\aaO_{i+1} \aO_{i} 
    +\text{h.a.}) 
    + \frac{V}{2} \sum_{i=1}^{I} \nO_i (\nO_i - 1)
}
with a modified hopping term that contains the so-called Peierls phase
factors $\ee^{\pm\ii\, \Theta/I}$ \cite{ShSu90}. The energy $E_{\Theta}$
in the presence of the phase twist is given by the lowest eigenvalue of
the twisted Hamiltonian using periodic boundary conditions. From the
difference of the ground state energies $E_{\Theta}-E_0$ we obtain the
superfluid fraction
\cite{RaSc99}
\eq{ \label{eq:sffraction}
  f_{\SF}
  = \frac{I^2}{N}\; \frac{E_{\Theta}-E_{0}}{J\,\Theta^2} \;,
}     
now expressed in terms of the parameters of the Bose-Hubbard model.


One can get a more detailed insight into the dependency of the
superfluid fraction on the structure of the eigenstates of the system by
considering a perturbative calculation of the energy difference
$E_{\Theta}-E_0$. We expand the the twisted Hamiltonian up to second order
in the twist angle $\Theta$ thus,
\eq{ 
  \HO_{\Theta}
  \approx \HO_0 + \frac{\Theta}{I}\,\JO - \frac{\Theta^2}{2 I^2}\,\TO
  = \HO_0 + \HO_{\text{pert}} \;.
}
Here we defined a current operator $\JO = \ii J \sum_i (\aaO_{i+1} \aO_i
- \text{h.a.})$ and the usual kinetic energy or hopping operator $\TO =
-J \sum_i (\aaO_{i+1} \aO_i + \text{h.a.})$. We can calculate the energy
shift $E_{\Theta}-E_{0}$ caused by the perturbation $\HO_{\text{pert}}$
in second order perturbation theory.  Retaining the terms up to the
quadratic order in the twist angle $\Theta$ we obtain for the superfluid
fraction using Eq. \eqref{eq:sffraction}
\eqmulti{ \label{eq:sffrac_drude}
  f_{\SF} 
  &= f_{\SF}^{(1)} - f_{\SF}^{(2)} \\
  &= \frac{1}{NJ} \bigg(\! -\frac{1}{2}\matrixe{\Psi_0}{\TO}{\Psi_0}
    - \sum_{\nu\ne0} \frac{|\matrixe{\Psi_{\nu}}{\JO}{\Psi_0}|^2}{E_{\nu}-E_{0}}
    \bigg) ,
}
where the $\ket{\Psi_{\nu}}$ ($\nu=0,1...$) are the eigenstates of  the
non-twisted Bose-Hubbard Hamiltonian $\HO_0$. The ground state expectation
value of the hopping operator describes the first order contribution
$f_{\SF}{}^{(1)}$. The sum over the excited states involving the current
operator constitutes the second order term $f_{\SF}{}^{(2)}$. It is this
second order term which introduces a significant dependence of the
superfluid fraction on the excitation spectrum and thus goes beyond the
static ground state properties of the system.

Equation \eqref{eq:sffrac_drude} corresponds to the Drude weight used to
characterize the DC conductivity of charged fermionic systems
\cite{FyMa91}. This demonstrates that the phase factors appearing in the
twisted Hamiltonian \eqref{eq:hamiltonian_twist} can actually be realized
experimentally, i.e., by an external electric field for charged particles
or by some linear external potential or even by accelerating the lattice.

\begin{figure}
  \includegraphics[width=0.8\columnwidth]{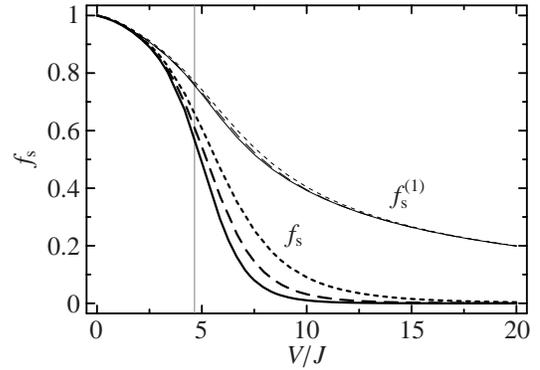}
  \vspace*{-3ex}
  \caption{Superfluid fraction $f_{\SF}$ as function of the interaction
  strength $V/J$ for filling $N/I=1$ and different lattice sizes: $I=10$
  (full line), $8$ (dashed), $6$ (dotted). The thin lines show the first
  order term $f_{\SF}^{(1)}$. The vertical
  line marks the critical interaction strength
  $(V/J)_{\text{crit}}\approx4.65$ for the infinite system.}
  \label{fig:superfluid} 
\end{figure}  

To illustrate the generic behavior of the superfluid fraction as function
of the interaction parameter $V/J$, and hence the appearance of the Mott
insulator phase, we solve the eigenvalue problems of the non-twisted
Hamiltonian \eqref{eq:hamiltonian} and the twisted Hamiltonian
\eqref{eq:hamiltonian_twist} numerically. We construct the corresponding
Hamilton matrices using a complete basis of Fock states
$\ket{n_1,...,n_I}$ with all compositions of the occupation numbers $n_i$.
The ground state energies $E_0$ and $E_{\Theta=0.1}$ are obtained with an
efficient iterative Lanczos algorithm \cite{RoBu02}. The perturbative
expression \eqref{eq:sffrac_drude} is then used to separate the
contributions of ground and excited states.

Figure \ref{fig:superfluid} shows the superfluid fraction $f_{\SF}$ for
one-dimensional lattices with up to $I=10$ sites and fixed filling factor
$N/I=1$. The superfluid fraction is $1$ for the noninteracting system and
decreases slowly for small $V/J$. In the region of the Mott transition
$f_{\SF}$ drops rapidly and goes to zero in the Mott-insulator phase. The
sequence of curves for increasing system size shows a moderate size
dependence around the onset of the insulator phase. One can extrapolate
the curves to $I\to\infty$ and finds a vanishing superfluid fraction above
a critical interaction strength which is in good agreement with the value
$(V/J)_{\text{crit}}\approx4.65$ obtained by strong coupling expansion
\cite{FrMo96} and Monte Carlo methods \cite{BaSc92}. The thin lines in
Fig. \ref{fig:superfluid} depict the isolated first order contribution
$f_{\SF}{}^{(1)} = -\frac{1}{2NJ}\matrixe{\Psi_0}{\TO}{\Psi_0}$ to the
superfluid fraction \eqref{eq:sffrac_drude}, which is just the reduced
expectation value of the hopping operator. This quantity decreases much
slower than the total superfluid fraction. Thus even deep in the Mott
regime the hopping operator has a considerable expectation value (up to
30\% of its value in the non-interacting system) although the system is
already an insulator, i.e., the superfluid fraction is zero. The rapid
decrease of the total superfluid fraction is largely due to the second
order contribution $f_{\SF}{}^{(2)}$, which vanishes for small $V/J$ and
shows a threshold-like behavior around the Mott transition. The vanishing
of the superfluid fraction in the insulating phase is generated by a
strong cancellation between the first and second order term. This
emphasizes that the superfluid fraction depends crucially on the properties
of the excited states.

\paragraph*{Interference Pattern.}

The standard experimental approach to investigate the state of the Bose
gas in an optical lattice relies on the matter-wave interference pattern
after the gas was released from the lattice. How much can the presence
or absence of interference fringes tell about superfluidity?

In the simplest model of the expansion for the system after release from
the lattice the effects of interactions are neglected. We can write the
intensity observed after some time-of-flight $\tau$ at a point $\yV$ as
\eq{ \label{eq:intensity1}
  \mathcal{I}(\yV)
  = \matrixe{\Psi_0}{\AAO(\yV)\AO(\yV)}{\Psi_0} \;.
}
We assume that the Wannier functions $w(x-\xi_i)$ can be described
by Gaussians. The amplitude operator is given by $\AO(\yV) =
\frac{1}{\sqrt{I}}\sum_{i=1}^{I} \chi_i(\yV)\; \aO_i$, where $\chi_i(\yV)$
denotes the Gaussian wave packet associated with site $i$ after a free
evolution for a time $\tau$. Since we are interested only in the generic
features of the interference pattern we discard all terms related to
the spatial structure of the envelope of $\chi_i(\yV)$ and only retain
the phase terms. This leads to
\eq{ \label{eq:amplitudeop}
  \AO(\yV)
  = \frac{1}{\sqrt{I}} \sum_{i=1}^{I} \ee^{\ii\, \phi_i(\yV)}\; \aO_i \;,
}
where $\phi_i(\yV)$ is the total phase acquired on the path from site $i$
to the observation point $\yV$. In the far-field approximation we can
assume a constant phase difference $\delta\phi(\yV) = \phi_{i+1}(\yV) -
\phi_{i}(\yV)$ for adjacent sites (gravity neglected). Calculating the
intensity \eqref{eq:intensity1} as function of the phase difference
$\delta\phi$, using the far-field form of the amplitude operator
\eqref{eq:amplitudeop}, leads to the following expression:
\eqmulti{ \label{eq:intensity}
  \mathcal{I}(\delta\phi)
  &= \frac{1}{I} \sum_{i,j=1}^{I}
    \ee^{\ii\,(j-i) \delta\phi}\; \matrixe{\Psi_0}{\aaO_i\aO_j}{\Psi_0} \\
  &= \frac{1}{I} \bigg[N 
    + \sum_{d=1}^{I-1} B_d\;\cos(d\,\delta\phi)\bigg] \;.
}
In the last step we rearranged the double summation into a sum over the
hopping distance $d=j-i$. The coefficients $B_d$ are given by the
expectation values of the $d$th neighbour hopping operators
\eq{ \label{eq:hoppingexpect}
  B_{d} 
  = \sum_{i=1}^{I-d}
    \matrixe{\Psi_0}{\aaO_{i+d} \aO_i + \aaO_i \aO_{i+d}}{\Psi_0} \;.
}
The reader should note that the leading coefficient $B_1$ is related to
the first order term of to the superfluid fraction \eqref{eq:sffrac_drude}
through $B_1 = 2(I-d)\, f_{\SF}{}^{(1)}$. Clearly, there is no
contribution corresponding to the important second order term
$f_{\SF}{}^{(2)}$ of the superfluid fraction, because the intensity
\eqref{eq:intensity1} involves only the ground state. Thus the
interference pattern cannot provide full information on the superfluid
properties, as it only measures the first order term of the superfluid
fraction.

\begin{figure}
  \includegraphics[width=0.8\columnwidth]{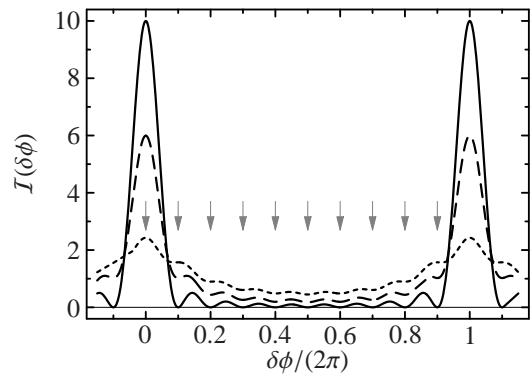}
  \vspace*{-2ex}
  \caption{Intensity $\mathcal{I}(\delta\phi)$ as
  function of the phase difference $\delta\phi$ for a system with $I=N=10$
  and $V/J=0$ (full line), $5$ (dashed line), $10$ (dotted line).}
  \label{fig:interference} 
\end{figure}  

Figure \ref{fig:interference} shows the intensities
$\mathcal{I}(\delta\phi)$ resulting from the exact numerical calculation
for different interaction strengths. The interference peaks around
$\delta\phi=0,\pm2\pi,...$ correspond to the prominent peaks observed
experimentally. Recall that terms describing the overall envelope were
neglected in \eqref{eq:amplitudeop}. With increasing $V/J$ the intensity
$\mathcal{I}_{\max}$  of these principal peaks reduces. At the same time
an incoherent background emerges such that the minimum intensity
$\mathcal{I}_{\min}$ between the principal peaks grows. Thus the
interference fringes are increasingly washed-out and eventually only a
flat intensity distribution remains. 

\begin{figure}
  \includegraphics[width=0.8\columnwidth]{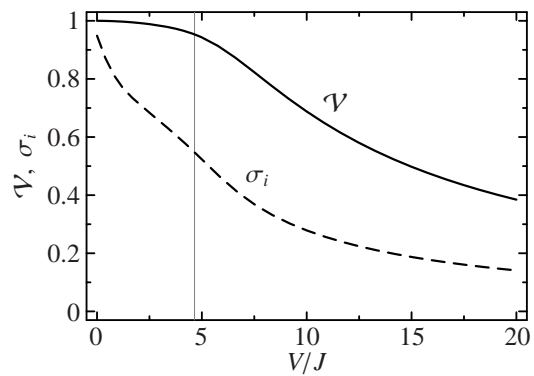}
  \vspace*{-2ex}
  \caption{Visibility $\mathcal{V}$ of the interference fringes
  (full line) and number fluctuations $\sigma$ (dashed) as function of the
  interaction strength $V/J$ for a system with $I=N=10$. }
  \label{fig:visibility} 
\end{figure}  

As a quantitative measure for the vanishing of the interference pattern
the full line in Fig. \ref{fig:visibility} shows the fringe visibility
$\mathcal{V} =
(\mathcal{I}_{\max}-\mathcal{I}_{\min})/(\mathcal{I}_{\max}+\mathcal{I}_{\min})$
as function of $V/J$. In addition, the dashed curve shows the on-site
number fluctuations $\sigma_i = (\expect{\nO_i^2} -
\expect{\nO_i}^2)^{1/2}$ of the ground state. Obviously, the visibility of
the fringes has no immediate relation to the number fluctuations. For
small interaction strengths $V/J\lesssim5$ the visibility remains almost
constant at $\mathcal{V}\approx 1$ whereas the number fluctuations drop to
$0.5$ in the same interval. 

A second observation concerns the relation with the superfluid fraction
shown in Fig. \ref{fig:superfluid}. The superfluid fraction $f_{\SF}$
vanishes much faster than the visibility $\mathcal{V}$ and the number
fluctuations $\sigma_i$. For values of $V/J$ where the superfluid fraction
has practically vanished the visibility is still larger than $0.7$. Thus
neither fringe visibility nor number fluctuations are a suitable
indicator for the superfluid properties and the Mott transition in lattice
systems. 

\enlargethispage{6ex}
\paragraph*{Quasi-Momentum Distribution.}

The interference pattern after ballistic expansion is closely related to
the quasi-momentum distribution of the Bose gas in the lattice. Formally,
the connection is revealed by constructing an expression for the
occupation numbers for the Bloch states of the lowest band. We can use the
relation between localized Wannier functions $w(x-\xi_i)$ and Bloch
functions $\psi_q(x)$ to define a creation operator $\ccO_q$ for a boson
in Bloch state with quasi-momentum $q$ \cite{OoSt01}
\eq{ \label{eq:bloch_wannier_op}
  \ccO_q 
  = \frac{1}{\sqrt{I}} \sum_{i=1}^{I} \ee^{-\ii\,q\,\xi_i}\; \aaO_i \;,
}
where $\xi_i$ is the coordinate of the center of the $i$th lattice site.
This relation is identical to the definition of the amplitude operator
$\AO(\yV)$ in \eqref{eq:amplitudeop} if we identify the phase $\phi_i(\yV)$
with $q\xi_i$ or the phase difference $\delta\phi$ with $qa$, where $a =
\xi_{i+1}-\xi_i$ is the lattice spacing. The occupation numbers
$\tilde{n}_q$ for the Bloch states with quasi-momenta $q$ are, therefore,
directly related to the intensity \eqref{eq:intensity} through
\eq{ \label{eq:momentumdist}
  \tilde{n}_q 
  = \matrixe{\Psi_0}{\ccO_q\cO_q}{\Psi_0} 
  = \mathcal{I}(\delta\phi=q a) \;.
}
Notice that in a finite system of length $L$ the quasi-momentum $q$ has
discrete values which are integer multiples of $2\pi/L$. The values of
$\delta\phi=qa$ for these allowed quasi-momenta are marked by gray arrows
in Fig. \ref{fig:interference}.

Because of this intimate relation the interference pattern provides
complete information on the quasi-momentum distribution of the trapped
system. The intensity of the principal interference peak is proportional
to the occupation number of the $q=0$ Bloch state, i.e., it describes the
number of particles in the condensate. The washing out of the interference
peaks with increasing interaction strength is linked to the successive
redistribution of the population from the condensate state with $q=0$ to
states of higher quasi-momenta. In the limit of large $V/J$ the intensity
distribution is flat, i.e., all quasi-momentum states of the lowest band
are occupied uniformly.  On the basis of this one-to-one correspondence
between interference pattern and quasi-momentum distribution we can
reinterpret the visibility $\mathcal{V}$ of the interference fringes as
measure for the uniformity of the quasi-momentum distribution. Vanishing
visibility corresponds to a completely uniform quasi-momentum
distribution, whereas visibility $\mathcal{V}=1$ means that at least one
quasi-momentum state is unoccupied.

\paragraph*{Conclusions.}

We have shown that the matter-wave interference pattern observed
experimentally contains all the information on the quasi-momentum
distribution of the lattice system but no direct information on the
superfluid fraction. The behavior of the superfluid fraction shown in Fig.
\ref{fig:superfluid} depends strongly on the properties of the excitation
spectrum, which enters through the second order contribution
$f_{\SF}{}^{(2)}$. The importance of this second order term shows that one
cannot probe superfluidity through quantities which are only sensitive to
the ground state of the system (like number fluctuation, condensate
fraction, coherence properties, etc.). One has to devise an experimental
scheme that measures superfluidity directly. The formal definition of
superfluidity gives a hint how to accomplish this. As mentioned earlier
there are several methods to create the phase factor appearing in
$\HO_{\Theta}$ experimentally, e.g. by accelerating the lattice. By
observing the resulting flow behavior after release from the lattice one
should be able to distinguish superfluid and non-superfluid components and
determine the superfluid fraction.

This work was supported by the Deutsche Forschungsgemeinschaft, the UK
EPSRC, and the EU under the Cold Quantum Gases Network.


\end{document}